\documentclass[a4paper, 11pt]{article}
\usepackage{apacite}
\usepackage{natbib}
\usepackage[dvipsnames]{xcolor}
\usepackage{amsmath}
\usepackage{amsmath}
\usepackage{subfigure}
\usepackage{graphicx}
\usepackage{url,booktabs,multirow}
\usepackage{hhline}
\usepackage{amsmath,amsfonts,amssymb,amsbsy,amsthm,bm}
\usepackage{nicefrac}
\usepackage[normalem]{ulem}
\usepackage{subfigure}
\usepackage{color}
\usepackage{times}
\usepackage{comment}
\usepackage{setspace}
\usepackage[bottom]{footmisc}
\usepackage{caption}
\usepackage[flushleft]{threeparttable}
\graphicspath{ {./Figures/} }
\usepackage{lineno,hyperref,amsmath}

\usepackage{xparse}
\usepackage{amsmath}

\NewDocumentCommand{\tens}{t_}
 {%
  \IfBooleanTF{#1}
   {\tensop}
   {\otimes}%
 }
\NewDocumentCommand{\tensop}{m}
 {%
  \mathbin{\mathop{\otimes}\displaylimits_{#1}}%
 }

\title{Small area estimation for composite indicators: the case of multidimensional poverty incidence}
\author
{Alejandra Arias-Salazar$^{1}$, Andrés Gutiérrez$^{2}$, Xavier Mancero$^{2}$, \\ Stalyn Guerrero-Gómez $^{2}$,  Natalia Rojas-Perilla $^{3}$,   Hanwen Zhang$^{4}$ \\
\\
\normalsize{$^1$ University of Costa Rica },\\
\normalsize{$^2$ Economic Commission for Latin America and the Caribbean}, \\
\normalsize{$^3$ United Arab Emirates University}, 
\normalsize{$^4$ Universidad Autónoma de Chile}
}

\date{}
\begin{document}
\maketitle

\begin{abstract}\small

This paper proposes a methodology to obtain estimates in small domains when the target is a composite indicator. These indicators are of utmost importance for studying multidimensional phenomena, but little research has been done on how to obtain estimates of these indicators under the small area context. Composite indicators are particularly complex for this purpose since their construction requires different data sources, aggregation procedures, and weighting which makes challenging not only the estimation for small domains but also obtaining uncertainty measures. As case study of our proposal, we estimate the incidence of multidimensional poverty at the municipality level in Colombia. Furthermore, we provide uncertainty measures based on a parametric bootstrap algorithm.

\end{abstract}


\clearpage

\section{Introduction}\label{sec:intro}

Official statistics are a useful tool for decision-makers, as they provide information on the characteristics of a country's population and allow them to apply and monitor public policies aimed at specific population groups.
Composite indicators are commonly used to summarize complex phenomena that consider two or more dimensions. These types of indicators are implemented in different areas, e.g., social science: Human Development Index (HDI) \citep{UNDP23}, Gender Development Index \citep{UNDP23GDI}, Multidimensional Poverty Index \citep{UNDP23MPI}, Corruption Perceptions Index \citep{transparency2023}, environment: Environmental Performance Index \cite{EPI2022}, 
 economy: Composite of Leading Indicators \citep{OECDCLI2023}.

A composite indicator facilitates a joint analysis of relevant aspects, allowing for a more comprehensive understanding of complex phenomena, enables better communication of the results, and facilitates decision-making \citep{CEPAL2013}. 

Besides their advantages for public policy, the construction and use of composite indicators should be handled carefully. For example, the application of different weighting and aggregation methods could lead to misleading results, and the quality of the different data sources used for their construction should be evaluated \citep{freudenberg2003composite}. 

Among the challenges in the construction of composite indicators and their applicability is the disaggregation. The demand for disaggregated information is quickly increasing, since it is essential to develop and implement strategies to improve the quality of life of the inhabitants of a country, on issues such as employment, poverty, education, and health. A recent case that properly illustrates this need are the Sustainable Development Goals, which seek to provide information disaggregated by different relevant characteristics of the population \citep{UnitedNationsGeneralAssembly2015ResDevelopment}.

Small area estimation (SAE) has been proposed as a solution to the dissaggregation problem. SAE methods have the goal of producing accurate estimates in small domains with adequate precision, by combining two or more sources of information. Most of these methodologies, usually supported by unit or area-level regression models, provide efficiency gains if the correlation between existing auxiliary information and the survey data is sufficient \citep{Pfeffermann2013, Rao2015,  Pratesi2016, tzavidis2018start}. However, to the best of our knowledge, no literature has been produced on obtaining quality estimates of composite indicators within the small area context. The literature related to this subject is mainly focused on dimensionality reduction for latent indicators such as economic well-being (see e.g., \cite{moretti2020multivariate} and \cite{moretti2021small}). 

Although the construction of composite indicators requires some standard steps \citep{joint2008handbook}, we focus only on the estimation of indicators that are already defined well as their weights. That means we do not deal with dimensionality reduction, normalization, aggregation, and weighting aspects.  

Composite indicators can use different ways to aggregate the information. One option is to first aggregate at the population level and then combine dimensions. An example for this is the Human Development Index, which is based on summary indicators for the country. A second option is to first aggregate across dimensions and then across individuals. This is the case of indices such as the Unmet Basic Needs index or the Multidimensional Poverty Index \cite{Alkire2007}. Here we present a proposal for a composite indicator of the latter type, which first aggregates indicators defined as dichotomous variables at the individual level.

To illustrate our proposal, we take the incidence of multidimensional poverty as example. 
The index is based on the \cite{Alkire2007} methodology, which identifies deprivations for each relevant dimension of well-being and then aggregates them at the household level. The main difference is that the index presented here corresponds only to the incidence, which is a component of the Multidimensional Poverty Index. The incidence of multidimensional poverty that we define here is only for illustration purposes.

The components of a multidimensional poverty index are usually estimated using data from household surveys. These instruments usually collect information on many different dimensions of well-being at the household level, which is required by the \cite{alkire2010} method and other approaches that consider the overlap of simultaneous deprivations. Surveys in Latin America have also the benefits of being frequently updated (on a yearly basis in most countries) and having nationally representative samples. 

One of the shortcomings of household surveys is their limited ability for disaggregation for specific groups of population and geographical areas. In addition to the poverty rate at the national level, it is desirable to identify which groups of the population are more afflicted by poverty and what is the relative contribution of each particular deprivation to their poverty level, so as to provide useful information for the implementation of public policies.

Data sources such as administrative records and population censuses are better suited for attaining higher levels of disaggregation, but they face their own particular limitations. In Latin American countries, administrative records are usually not accessible at the individual level, do not have the necessary quality to produce reliable statistics, or do not provide information on different deprivations for the same individuals. On the other hand, population censuses are usually produced only every 10 years and they collect information on a restricted set of variables, that do not allow the calculation of a complete multidimensional poverty index.

The objective of this paper is to provide a methodology to produce disaggregated estimations at the municipal level of the incidence of a multidimensional poverty index using small-area methods, taking Colombia as a case study. To that end, model-based estimation methods are applied to integrate data from the survey with the population census. The methodology aims to preserve the information of each deprivation indicator so that the final index (incidence of multidimensional poverty) can be decomposed by dimension.  As an uncertainty measure, the mean squared error (MSE) is derived via parametric bootstrap.

The structure of the paper is as follows: the proposal to obtain small area estimates for composite indicators is explained in Section \ref{sec:methodology}, as well as the procedure to obtain MSE estimates for the corresponding point estimates. We present a case study to show the implementation of our proposal in Section \ref{sec:case_study}.
In Section \ref{sec:evaluation} we present a simulation exercise to validate our proposal. Conclusions and further research are presented in Section \ref{sec:conclusion}.

\section{Incidence of multidimensional poverty: a methodology for small area estimation}\label{sec:methodology}

Unlike other problems in SAE, getting a final indicator at the domain level (e.g., a total or mean) is not useful in this case. Because of its nature, the computation of the incidence of multidimensional poverty requires information on each individual for each of the indicators.
That means, the challenge is to first estimate the status of deprivation (deprived or not deprived) for each indicator across all persons in the census.

In general, the type of data required by the various composite indicators may differ. Some indicators are constructed from numerical variables, for example, life expectancy, income or years of schooling. In this paper, we focus on presenting a proposal to obtain small area estimates for a composite indicator consisting only of dichotomous variables. 

 In this section, we first explain the example that we use in our case study: the incidence of multidimensional poverty. Next, we introduce small area estimation methods for binary response variables (indicators). Then, we present our proposals where more than one indicators are missing in the census. Finally, we also describe how we address the problem of finding an uncertainty measure for this scenario.

\subsection{The incidence of multidimensional poverty}\label{subsec:mdi}

National statistical surveys often are not suitable to provide reliable socio-demographic estimates under small sample sizes at domain levels due to high costs. SAE procedures are estimation methodologies for obtaining such highly disaggregated target information under small sample sizes. Their basic principle is to improve classical procedures by combining survey and register data through a desired model.

In this paper, we focus on composite indicators that are constructed based on dichotomous variables. For this reason we selected as example the incidence of multidimensional poverty, since it is a well-known indicator. One particularity of \cite{Alkire2007} methodology where the Global Multidimensional Poverty Index is presented, is that each country can define its own set of indicators and dimensions based on its own necessities. Regardless of their specification,  multidimensional poverty indexes share the same characteristic: indicators explaining deprivations are grouped in within dimensions. 
Dimensions correspond to the relevant components of well-being that are related to the notion of poverty. The measurement of each dimension is operationalized through specific indicators, selected based on the information that is available, usually in household surveys. For each indicator, a deprivation cut-off is used to determine whether a person is to be considered deprived. 

Let us assume an index with $K$ indicators which are measured as deprivations: $y_{dj}^k=1$ if the person has the deprivation and $y_{dj}^k=0$ if the person has not had the deprivation. The index requires the information for each individual $j = 1, \dots , N_d$ in $d = 1, \dots, D$ domains, where $N_d$ denotes the population size of the domain $d$. The Global Multidimensional Poverty Index for the domain $d$ proposed by \citep{Alkire2007} is computed as:

\begin{equation*}
    \text{MDI}_d = \text{H} \cdot \text{A},
\end{equation*}

where H is the headcount ratio or incidence of multidimensional deprivations and A is the intensity. In this paper, we focus only on the incidence, i.e., the headcount ratio H, which is  defined as:

\begin{equation}\label{eq:mdi}
    \text{H}_d =\frac{1}{N_d} \sum^{N_d}_{j=1} I(q_{dj} > z).
\end{equation}

The indicator function $I(\cdot)$ equals 1 when the condition $q_{dj} > z$ is met, where $z$ is a thresholds defining if the person is poor or not.

As can be seen, to compute the incidence of multidimensional poverty (H), specifying if each individual has a deprivation (1) or not (0) is required. Therefore, the challenge is not to obtain ``final" indicator estimates (e.g. proportion of people with a specific deprivation by domain), as most SAE problems. In this case, the challenge is finding, if the person in the census has deprivation or not in the missing indicators.

\subsection{Small Area Estimation for Binary Variables}\label{subsec:sae}

In many applications the variable of interest in small areas can be binary, e.g., $y_{dj} =0$ or $1$ representing the absence (or not) of a specific characteristic. For a binary case, the target estimation in each domain $d=1, \dots, D$ can be the proportion $\bar{Y}_d = \pi_d = \frac{1}{N_d} \sum_{j=1}^{N_d} y_{dj} $ of the population having this characteristic, being $\pi_{dj}$ the probability that a specific unit $j$ in the domain $d$ obtains the value 1.

Although other methods have been proposed for binary outcomes, e.g.,  based on M-quantile modelling \citep{chambers2016semiparametric}, in this application we follow the traditional approach based on generalized linear mixed models. Under this scenario, the $\pi_{dj}$ with a logit link function defined as: 

    \begin{equation}\label{eq:Plugin1}
        \text{logit}(\pi_{dj}) = \text{log} \Big (   \frac{\pi_{dj}} {1-\pi_{dj}} \Big) = \eta_{dj} = \mathbf{x}^T_{dj}\beta + u_d
    \end{equation}
     
with $j=1, \dots, N_d$, $d=1, \dots, D$, $\beta$ a vector of fixed effect parameter, and $u_d$ the random area-specific effect for the domain $d$ with $u_d \sim N(0,\sigma^2)$. $u_d$ are assumed independent and $y_{dj}|u_d \sim \text{Bernoulli}(\pi_{dj})$ with $E(y_{dj}|u_d) = \pi_{dj}$ and $Var(y_{dj}|u_d) = \sigma^2_{dj}=\pi_{dj}(1-\pi_{dj})$. Furthermore, $\mathbf{x}_{dj}$ represents the $p\times 1$ vector of values of $p$ unit-level auxiliary variables. 

Since our specific problem is to find deprivations (0,1) for several indicators, we use this unit-level Bernoulli logit mixed model as the starting point. Derivations of different algorithms to fit the unit-level logit mixed model can be found in \cite{morales2021course}, namely: method of simulated moments (MSM), expectation-maximization (EM) algorithm, penalized quasi-likelihood (PQL) algorithm \citep{gonzalez2007estimation}, or maximum likelihood Laplace (ML - Laplace) approximation algorithm which is described in \cite{morales2021course}. For ease, we implement the latest algorithm as it is available in the  \texttt{R} \texttt{lme4} package.

The goal of obtaining quality estimates in small domains can be jeopardized when several of these domains are not considered in the sample or the sample size is not enough to produce reliable results. An empirical best predictor (EBP) can be defined for this purpose \citep{jiang2001empirical, jiang2003empirical}. 

The EBP for quantities of interests such as probabilities, sums of probabilities, and proportions by domains can be approximated using Monte Carlo simulation \citep{hobza2016empirical}. In practice, this option is usually avoided since it does not have a closed form requiring numerical approximation for it computation \citep{chambers2016semiparametric}. As a solution, the plug-in predictor of $\pi_{dj}$ is defined as:
       
       \begin{equation}\label{eq:plugin}
           \hat{\pi}^{in}_{dj} = \frac{\text{exp}(\mathbf{x}^{T}_{dj}\hat{\beta}+\hat{u}_d)} {{1+\text{exp}(\mathbf{x}^{T}_{dj}\hat{\beta}+\hat{u}_d)}},
       \end{equation}

which would allow obtaining the plug-in predictor of $\bar{Y}_d$:

\begin{equation}
    \hat{\bar{Y}}^{in}_{d} = \frac{1}{N_d} \big( \sum_{j \in s_d} y_{dj} + \sum_{j \in r_d} \hat{\pi}^{in}_{dj} \big),
\end{equation}

where $s$ and $r$ represent the in- and out-of-sample observations respectively.

\subsection{Point Estimation for the H Predictor}\label{subsec:point}

Let $K$ be the missing indicators for each individual $j$ in the census. The proposed procedure to estimate H is as follows:

\begin{enumerate}
    \item  Use the sample data to fit a unit-level Bernoulli logit mixed model for each indicator and estimate $\hat{\beta}^k$, $\hat{u}^k_d$, and finally, $\hat{\pi}^{in,k}_{dj}$, with $k=1,\cdots,K$ as in Equation \ref{eq:plugin}.
    
\item For $l = 1, \dots, L$ Monte Carlo simulations:

\begin{itemize}

    \item For each individual in the census, predict the probability of obtaining the value 1 for the $k$-th indicator. i.e. $\hat{\pi}^{in,k, (l)}_{dj} \hspace{0.1cm} \forall \hspace{0.1cm} j \in U_d$.
    \vspace{0.2cm}
    
    \item Obtain Monte Carlo estimates $\tilde{y}^{k, (l)}_{dj}$ with $y^k_{dj}  \sim \text{Bernoulli}(\hat{\pi}^{in,k}_{dj})$.
    \vspace{0.2cm}
    
\item Compute the $\text{H}^{(l)}$, with the indicators already available in the census and the new indicators $\tilde{y}^{k,(l)}_{dj}$, as indicated in Equation \ref{eq:mdi}. 

\end{itemize}

\item The final point estimate in each small area $d$ is computed by taking the mean over each $L$ simulation: $$\widehat{\text{H}}_d = \frac{1}{L} \sum^L_{l=1}\text{H}^{(l)}_d.$$

\end{enumerate}


Note that under this proposal, the incidence of multidimensional poverty (H) can be estimated even if there are several missing indicators $K \geq 1$.

The proof of this approach can be found in Appendix \ref{proof:one} when only indicator is missing and in Appendix \ref{proof:two}, when two indicators are missing in the census.

\subsection{Estimation of the MSE}\label{subsec:MSE}

The estimation of the mean squared error (MSE) as the accuracy measure when using small area estimators is a key step when estimating socio-demographic information. In case the variable of interest is binary, some approximations are available for obtaining the analytic form of the MSE. \cite{gonzalez2007estimation} derived a  small area robust bootstrap (SAWB) for the uncertainty estimation of an empirical predictor. Based on this bootstrap scheme, we present a modification that allows to consider that the target estimate, i.e., the multidimensional poverty incidence, has several components or indicators and one or more of these indicators are estimated via SAE methods. 

The steps of the proposed parametric bootstrap are as follows: For each missing indicator, $Y_k$, with $k = 1,...,K$ and for $b = 1,...,B$, with $B$ denoting the number of bootstraps: 

	\begin{enumerate}
		\item Using the already estimated $\hat{\beta}^{k}, \hat{\sigma}^{2,k}_u$ as described in Section \ref{subsec:sae}, generate $u_{d}^{*,k}  \sim N(0, \hat{\sigma}^{2,k}_u)$ \textit{iid}.
		\item Simulate a bootstrap superpopulation for each indicator $y^{k,(b)}_{dj}  \sim \text{Bernoulli}({\pi}^{in,*,k}_{dj})$ with $   {\pi}^{in,*,k}_{dj} = \frac{\text{exp}(\mathbf{x}^{T}_{dj}\hat{\beta}^{k}+{u}_{d}^{*, k})} {{1+\text{exp}(\mathbf{x}^{T}_{dj}\hat{\beta}^{k}+{u}_{d}^{*, k})}}.$
		
		\item Calculate the $\text{H}_{d}^{(b)}$ as indicated in Equation \ref{eq:mdi}.

			\vspace{0.3cm}
			
			\item Extract the bootstrap sample and obtain the $\widehat{\text{H}}_{d}^{(b)}$ following the point estimate - Monte Carlo approach described in Section \ref{subsec:sae}.
		\vspace{0.2cm}

		\item	
		$
		\widehat{\text{MSE}}\Big[ \widehat{\text{H}}_{d} \Big]=1/B \sum\limits_{b=1}^{B}\Big[\text{H}_{d}^{(b)}-\widehat{\text{H}}_{d}^{(b)}\Big]^2.
		$		
	\end{enumerate}

\section{Case Study: multidimensional poverty incidence for the adult population in Colombia}\label{sec:case_study}

The case study presented in this paper uses an example of the incidence of a multidimensional poverty index (described in Equation \ref{eq:mdi}).  We use this index only for illustrative purposes. The composition of the index here presented (dimensions and indicators) is not the one defined by National Statistical Office of Colombia (in Spanish, Departamento Administrativo Nacional de Estadística, DANE). or another international organization.

For this example, we use data from Colombia, since it has a recent population and housing census (from 2018) and a household survey from the same year.

Table \ref{tab:mpi} describes the composition of the index that we use in this example. 
This index is based on previous research by UN-ECLAC on possible structures for a multidimensional poverty index that is comparable for Latin American countries, based on the availability of information from national household surveys \citep{CEPAL2014, santos2018multidimensional}. The index includes 5 dimensions (housing; water and sanitation; energy and connectivity; education; and employment and social protection) and $K=8$ indicators. Indicators and cut-offs are the same for adults and seniors, except in two cases. For education, the cut-off for insufficient education is 12 years for adults ages 18 to 29, nine years for adults ages 30 to 59, and four years for ages 60 and over. For the dimension of employment and social protection, the indicator “unemployment or insufficient employment income” for adults is replaced by “no pension or insufficient pension income” for seniors.

Data available in censuses usually includes the required information for calculating this index, except in the case of the employment and social protection dimensions, which require data on individual income. For the purpose of this paper, it has been assumed that information on education is also not available in the census and thus has to be estimated through SAE methods.

\begin{table}[ht]
	\begin{center} 	
	\caption{Composition of the index, availability of indicators
in the Colombian census and target population.}
\label{tab:mpi}
	\small{
			\centering
			\begin{tabular}{llccc}
				\hline
				\textbf{Dimension} & \textbf{Indicator}& \textbf{Weight} &\textbf{In census} &\textbf{Target}
					\\
				\hline
				Housing &Poor housing materials & 1/10 & Yes & Adults, seniors\\
				& Overcrowding & 1/10 &Yes & Adults, seniors \\
			\hline
				Water and 
				 & Lack of drinking water & 1/10 &Yes & Adults, seniors 
				\\
				sanitation & Lack of sanitation &1/10 & Yes & Adults, seniors \\
					\hline
					Energy and 
				  & 
				 Lack of internet service & 1/10 &Yes & Adults, seniors \\
				connectivity & Lack of electricity & 1/10 &Yes & Adults, seniors \\
				\hline 
				Education & Unfinished education & 2/10 &\textbf{No} &Adults, Seniors\\
				\hline
				Employment and  & No or insufficient pension & 2/10 & \textbf{No} & Seniors \\
			 social protection & Unemployment or insufficient & &  &Adults\\
			   &employment-related income  &&&\\
				\hline
		\end{tabular}
		}
		\end{center}

\end{table}

We focus on the incidence of multidimensional poverty described in Equation \ref{eq:mdi}. Here, we require $K=8$ indicators which are measured as deprivations: $y_{dj}^k=1$ if the person has the deprivation and $y_{dj}^k=0$ if the person has not had the deprivation. 

The index requires the information for each individual $j = 1, \dots , N_d$ in $d = 1, \dots, D$ domains, where $N_d$ denotes the population size of the domain $d$.

The indicator function $I(\cdot)$ equals 1 when the condition $q_{dj} > z$ is met. For the purpose of this paper, we use the value of 0.4 for $z$ i.e. $I(\cdot)$ equals 1 when $q_{dj} > 0.4$. $q_{dj}$ is a weighted quantity considering the $K=8$ indicators that comprise the index (see Table \ref{tab:mpi}):

$$ q_{dj} = 0.1  \sum^{6}_{k=1} y_{dj}^k + 0.2 \sum^{8}_{k=7} y_{dj}^k .$$

The first part of the sum considers the indicators for housing, water and sanitation, energy and connectivity dimensions. The second part, the indicators of education and employment and social protection dimensions. The latter are in fact the two missing indicators in the census and will be estimated with the methodology presented in Section \ref{subsec:point}.

\begin{figure}[!htb]
 \centering
\includegraphics[width=1.3\textwidth]{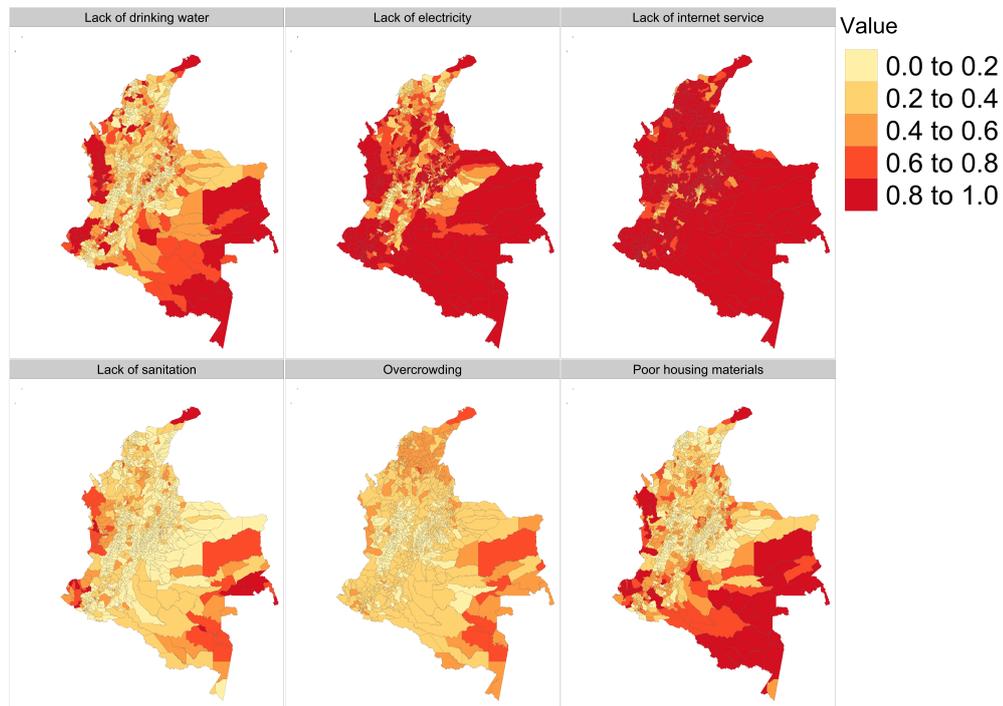}
\caption{Indicators of the H available in the census at the municipal level}\label{fig:indic_censo}
\end{figure}

Figure \ref{fig:indic_censo} shows the proportion of people who had deprivations in six of the eight indicators that make up H, i.e., the incidence of multidimensional poverty. These maps were generated at the municipal level with information from the census. The maps of the two missing indicators were not generated since the census does not include the required information. In addition, the calculation of the H is not possible. 

\subsection{Data Sources}\label{subsec:data}

\subsubsection{Censo Nacional de Población y Vivienda (CNPV)  2018}

The national population and housing census (in Spanish, Censo Nacional de Población y Vivienda, CNPV) is conducted by the DANE. Although it is planned to be every 10 years, the last census was carried out three years later as planned due to administrative and economic reasons. For first time, the information was collected through electronic self-interviewing, in addition to the traditional face-to-face interview. The data collection phase took place in 2018 for 10 months \citep{DANE2019}.

The census aims at collecting demographic information on the Colombian population and its living conditions, including housing and household characteristics. This information is essential for territorial planning and decision-making in the country \citep{DANE2019}.

\subsubsection{Gran Encuesta Integrada de Hogares (GEIH) 2018}

The Great Integrated Household Survey (in Spanish, Gran Encuesta Integrada de Hogares, GEIH) is conducted annually by the DANE. It provides information on the size and structure of the labor force, as well as the sociodemographic characteristics of the population and households, in addition to housing, educational level, affiliation to social security, income, among others (DANE, 2021). It is the official source of information for employment and income poverty indicators and includes the necessary variables to calculate the ECLAC multidimensional deprivation index. Administratively, Colombia is organized in 5 regions, 33 departments, and 1122 municipalities; the GEIH provides representative information at the national level, urban and rural areas, regions, and 24 of the 33 departments. In 2018, the valid survey sample included 231,128 households and 762,753 individuals.

The version of the GEIH used in this paper comes from the Household Survey Data Bank (BADEHOG), a repository of household surveys from 18 Latin American countries maintained by the ECLAC Statistics Division. In this repository, variables are harmonized to allow the construction of various indicators and their comparison across countries. This feature could simplify the estimation of the incidence of multidimensional poverty for the countries of the region.

\subsubsection{Satellite Imagery}

Geospatial data has become a powerful resource to improve quality in local-area estimations. Although big data, in general, has gained popularity for this purpose, geospatial data has two special advantages: it is easy to access and it is bias selection free  \citep{masaki2020small}.

Satellite imagery as auxiliary source of information has been implemented in several small area estimation problems in topics such as well-being \citep{engstrom2022poverty}, population density \citep{Harvey2002,Chengbin2013,Steinnocher2019}, poverty mapping \citep{Babenko2017,Chandra2018}, and multidimensional poverty \citep{pokhriyal2017combining,koebe2022intercensal}, among others.

For this case study, we make use of the resources available in the Earth Engine Data Catalog \citep{gorelick2017google}. Among many available products, we incorporate in our model area-level information on night light intensity, urban cover fraction, and crop cover fraction.

\subsection{Results}\label{sec:results}

In this Section, we present the main results of applying the proposed methodology to obtain estimates of the multidimensional poverty incidence described in \ref{subsec:mdi} in departments and municipalities of Colombia.

The distribution of sample and population sizes for the domains of interest is presented in Table \ref{tab:sizes}. The census size is $N=34,180,812$ and the sample has a size $n=549,077$ which covers 24 out of 33 departments and 438 out of 1122 municipalities.

\begin{table}[!ht] 	
	\small{
	\caption{Distribution of the sample and census sizes across departments and municipalities}\label{tab:sizes}
	\centering
		\begin{tabular}{llrrrrrr}
		\hline
         Domain&Source  &Min &1st Q &Median& Mean & 3rd Q & Max  \\ 
		\hline
		\textbf{Department} & && &&&&\\
			\hline
	 (In-sample:) & Survey & 6613 & 21601 & 22344 & 22878 & 24692& 35264   \\
	 73\%	  & Census &   21037 &256244 &752416 & 1035781 &1036702  &  5847519\\
			\hline
	\textbf{Municipality} &&& &&&&\\
				\hline
(In-sample: ) & Survey & 14 & 127 & 220 & 1254 &  369 & 24594 \\
	40\%	  & Census &  103 &4433  &8440 & 30464 &17601.25 & 5847519\\
		  		\hline
	\end{tabular}
	}
\end{table}

As can be seen in Table \ref{tab:sizes}, in this application the sample sizes are small only in some municipalities, and improving the accuracy of the estimates is in this case, not the main issue, as it is in many small area estimation problems. Nevertheless, we analyze the accuracy of the estimates based on the coefficients of variations, defined as

$$CV(\widehat{\text{H}}_{d})= \frac{\sqrt{MSE(\widehat{\text{H}}_{d}})}{\widehat{\text{H}}_{d}} \times 100.$$

\hspace{2cm}

Summary statistics of the CVs from the direct and model-based estimates are presented in Table \ref{tab:cvs}. Here, the CVs are also disaggregated by domain level (department and municipality), and for each missing indicator (education and employment). At the department level, the CVs are relatively small for both indicators and also for direct and model-based estimates. The uncertainty provided by the out-of-sample departments can explain the higher values for the model-based estimates. 
The benefit of using SAE methods is most apparent in municipalities since the average CVs are similar to the direct estimates but the third quartile and the maximum value are lower than the direct estimates. This behavior is observed for both indicators, education, and employment. 

\begin{figure}[!htb]
     \centering
	\includegraphics[width=15cm]{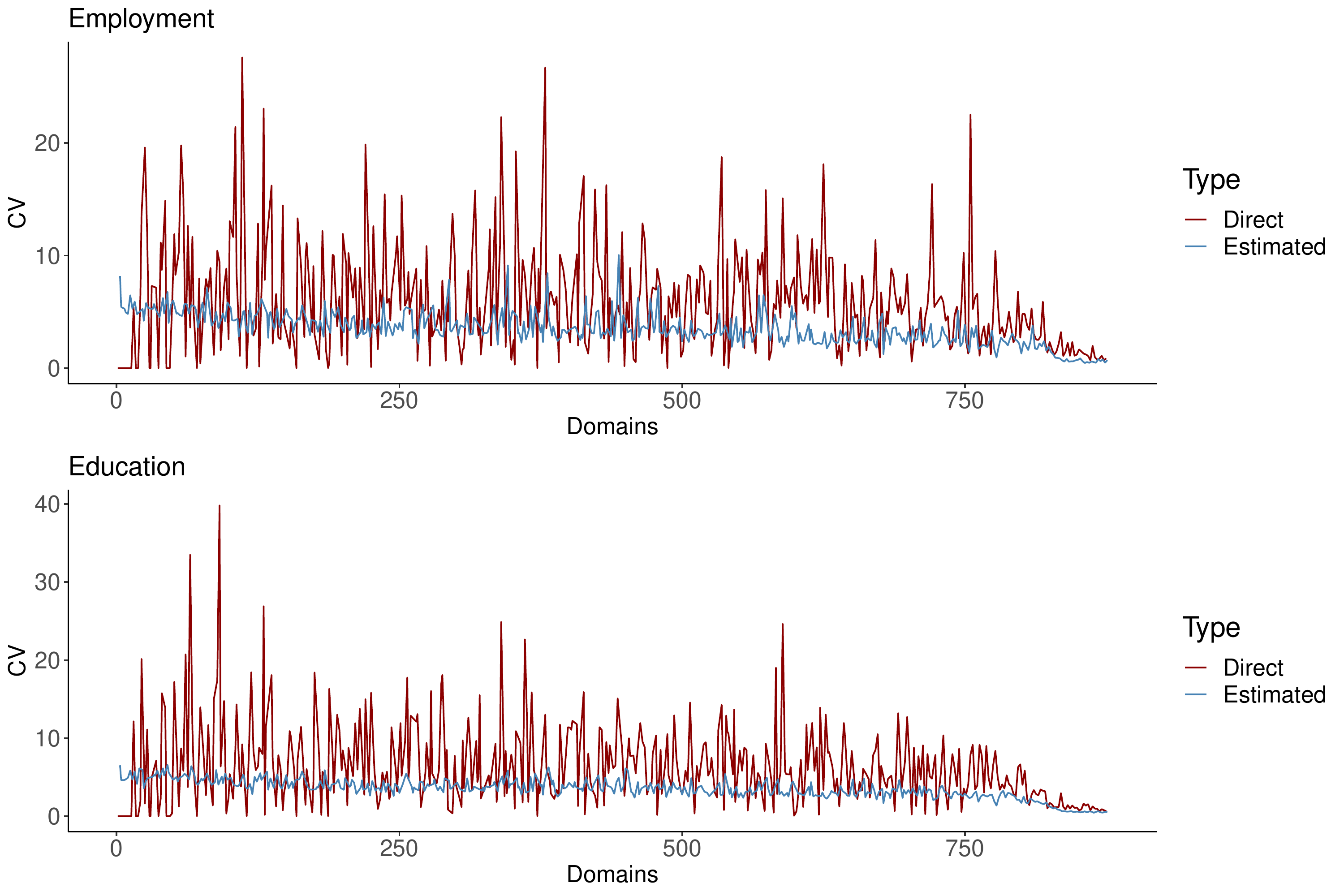}
 \caption{Coefficients of variation (in percentage) of the direct and model-based estimates at the department and municipality level for the indicators employment and education.}
 \label{fig:cvs}
 \end{figure}

The stable and low CVs that the model-based estimates provide, can be clearly observed in Figure \ref{fig:cvs}) for the in-sample municipalities. 

\begin{table}[!ht] 	
	\small{
	\caption{Descriptive statistics of the coefficients of variation across departments and municipalities (in percentage)}
	\centering
		\begin{tabular}{llrrrrrr}
		\hline
         Domain/Indicator& Estimation &Min &1st Q &Median& Mean & 3rd Q & Max  \\ 
		\hline
		\textbf{Department} & & &&&&\\
			\hline
	 Education & Direct & 0.64   & 1.24  &  1.43   & 1.47  &  1.67  &  2.09 \\

	  & Model-based & 0.36  &  0.44 &   0.49 &   1.56 &   2.85  &  9.22 \\
	  	\hline
	 Employment	  & Direct &  0.87 &   1.39  &  1.73  &  1.80   & 2.04  &  3.40\\
	   & Model-based &  0.26 &   0.40 &   0.53 &   1.66 &   2.90 &9.24\\
			\hline
	\textbf{Municipality} &&& &&&&\\
				\hline
 Education & Direct & 0.00  &  2.60  &  5.37   & 6.27  &  8.75 &  39.81\\

	  & Model-based &  0.45  &  3.81  &  5.37  &  5.23  &  6.51  & 10.87\\
	  	\hline
	 Employment	  & Direct &  0.00   & 2.33  &  5.19 &   5.91   & 8.46 &  27.60\\
	   & Model-based &  0.46 &   3.69  &  5.58 &   5.53  &  7.03  & 18.78\\
		  		\hline
	\end{tabular}
	}
\end{table}\label{tab:cvs}

\begin{figure}[!htb]
 \centering
\includegraphics[width=12cm]{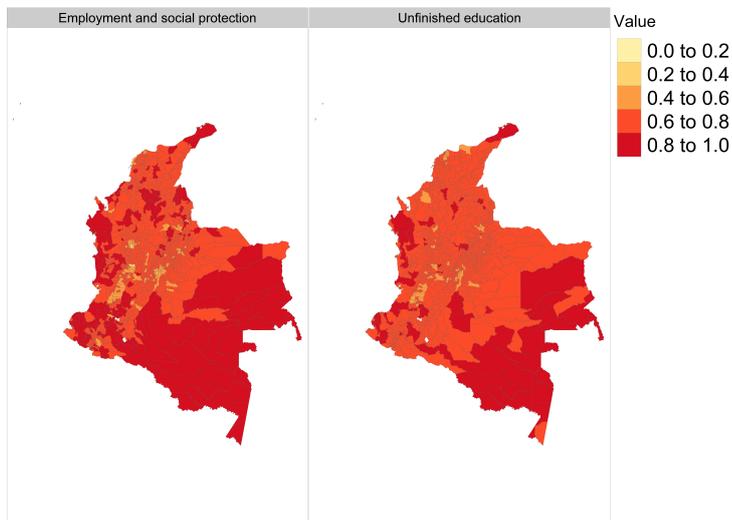}
\caption{Model-based estimates for the indicators employment and social protection and unfinished education at the municipality level.} 
\label{fig:modelbased_indicators}
\end{figure}

\begin{figure}[!htb]
 \centering
\includegraphics[width=12cm]{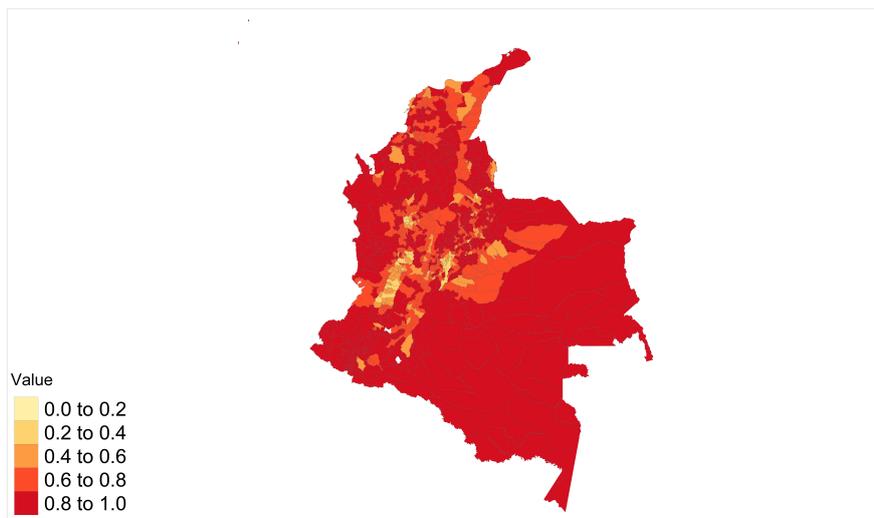}
\caption{Final H at the municipality level.} 
\label{fig:MDI_municip}
\end{figure}

\begin{figure}[!htb]
 \centering
\includegraphics[width=12cm]{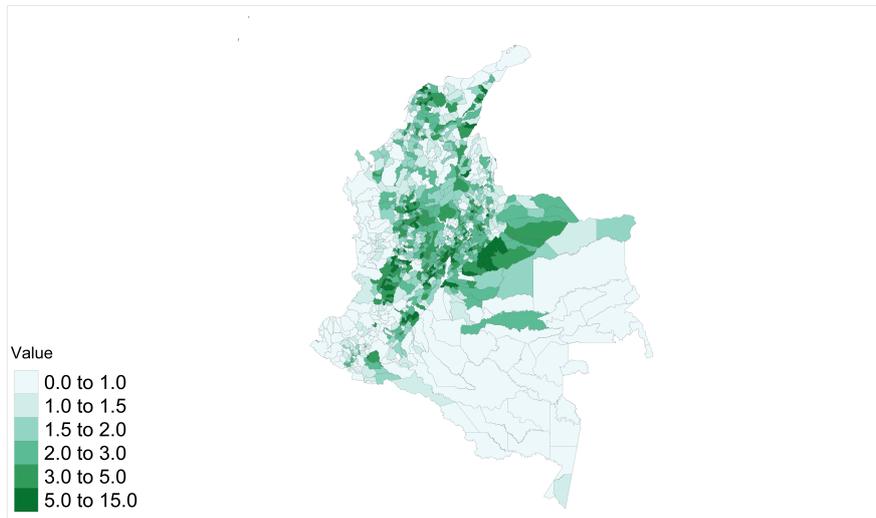}
\caption{Coefficients of variation of the H at the municipality level.} 
\label{fig:MDI_CV_municip}
\end{figure}

Since we observed that the CVs from direct estimates are acceptable for both indicators, the main benefit of applying the small area estimation method that we propose is to obtain reliable estimates for the out-of-sample domains: 9 departments and 684 municipalities, in order to provide the proportion of the population of interest under multidimensional poverty for all domains of interested. The results are showed in Figures \ref{fig:modelbased_indicators}-\ref{fig:MDI_CV_municip}.

Figure \ref{fig:modelbased_indicators} presents the two indicators at municipality level that were not available in the census data and required the use of SAE methods to obtain these estimates.

In general, Colombia is a country with diverse regions, each with its own unique set of economic and social challenges. As shown in Figure \ref{fig:modelbased_indicators}, one of the regions that requires special attention is the Amazon region, which encompasses departments such as Guainía, Vaupés, and Guaviare, as well as other departments such as Chocó, Guajira, and Sucre. These departments are characterized by high levels of poverty, low levels of education and employment, and limited access to basic services such as housing, water, and sanitation. In particular, education is a critical determinant of an individual's well-being and standard of living. Despite the recent progress made in increasing access to education in those departments, they tend to have lower enrollment rates and lower levels of educational attainment, resulting in many children not having access to quality education and facing a greater risk of poverty and exclusion from the formal economy. Employment and social protection is the second key indicator of poverty and well-being that was estimated. Even the unemployment rate in Colombia has decreased in recent years, there are still disparities in employment across the country. The departments mentioned above tend to have higher unemployment rates and a larger informal sector, resulting in many people being unable to find formal employment and facing a greater risk of poverty and exclusion from the formal economy. In some departments, such as Bogotá, there are higher levels of formal employment and educational attainment with a significant portion of the population working in the formal sector and having completed tertiary education.

With these two indicators, now it is possible to compute the final H, the result is shown in Figure \ref{fig:MDI_municip} for municipalities of Colombia. The CVs of the final H are below 15\% (Figure \ref{fig:MDI_CV_municip}), indicating that the estimates are ``acceptable" in terms of precision.

The estimates obtained with the proposed methodology allow deeper analysis from two points of view: What are the most severe deprivations? and what are the most affected areas in general? We take the department Valle del Cauca as an example, which is one of the departments with the lowest levels of the H (Figure \ref{fig:MDI_municip}: middle-west of the country with an MDI of 38.5\%).

\begin{figure}[!htb]
 \centering
\includegraphics[width=1\textwidth]{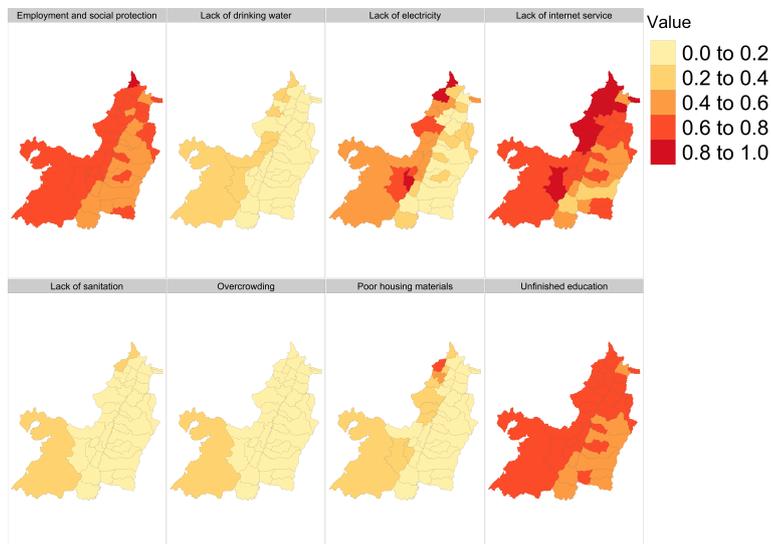}
\caption{Valle del Cauca: Indicator of the H at the municipality level.} 
\label{fig: VC_indicators}
\end{figure}

\begin{figure}[!htb]
 \centering
\includegraphics[width=12cm]{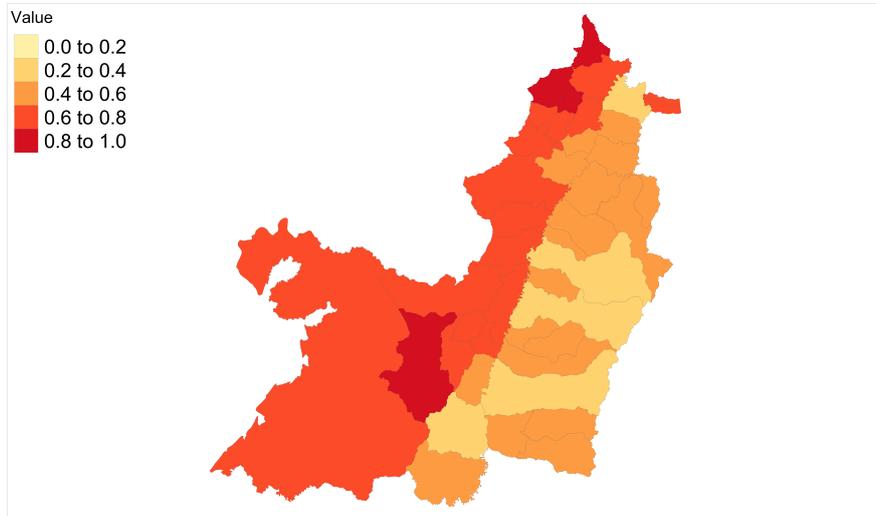}
\caption{ Valle del Cauca: final H at the municipality level.} 
\label{fig: VC_MDI}
\end{figure}

Valle del Cauca has a low proportion of the adult population with deprivations in housing, overcrowding, drinking water, sanitation, and electricity (between 0 and 20\%), and intermediate values (40-60\%) in internet service, education and employment, and health insurance (see Figure \ref{fig: VC_indicators}). Furthermore, when we take a look at the general H, it is possible to see that there are strong differences between municipalities in this department. Figure \ref{fig: VC_MDI} shows that the municipalities El Águila and Dagua present high levels of H (92\% and 82\% respectively), while other areas such as Cali and Tulua have relatively low values (30\% and 39\%). It is important to note that there are still significant disparities in poverty incidence across municipalities within the department. This underscores the need for targeted interventions to address these challenges and reduce multidimensional poverty in the department.

\section{Evaluation}\label{sec:evaluation}

The evaluation of the proposed method is twofold. First, an internal comparison between the direct and the model-based estimates is presented. Second, a design-based simulation study is described to evaluate the performance of the estimator proposed in Section \ref{subsec:sae}.\\

\begin{figure}[!htb]
 \centering
\includegraphics[width=11cm]{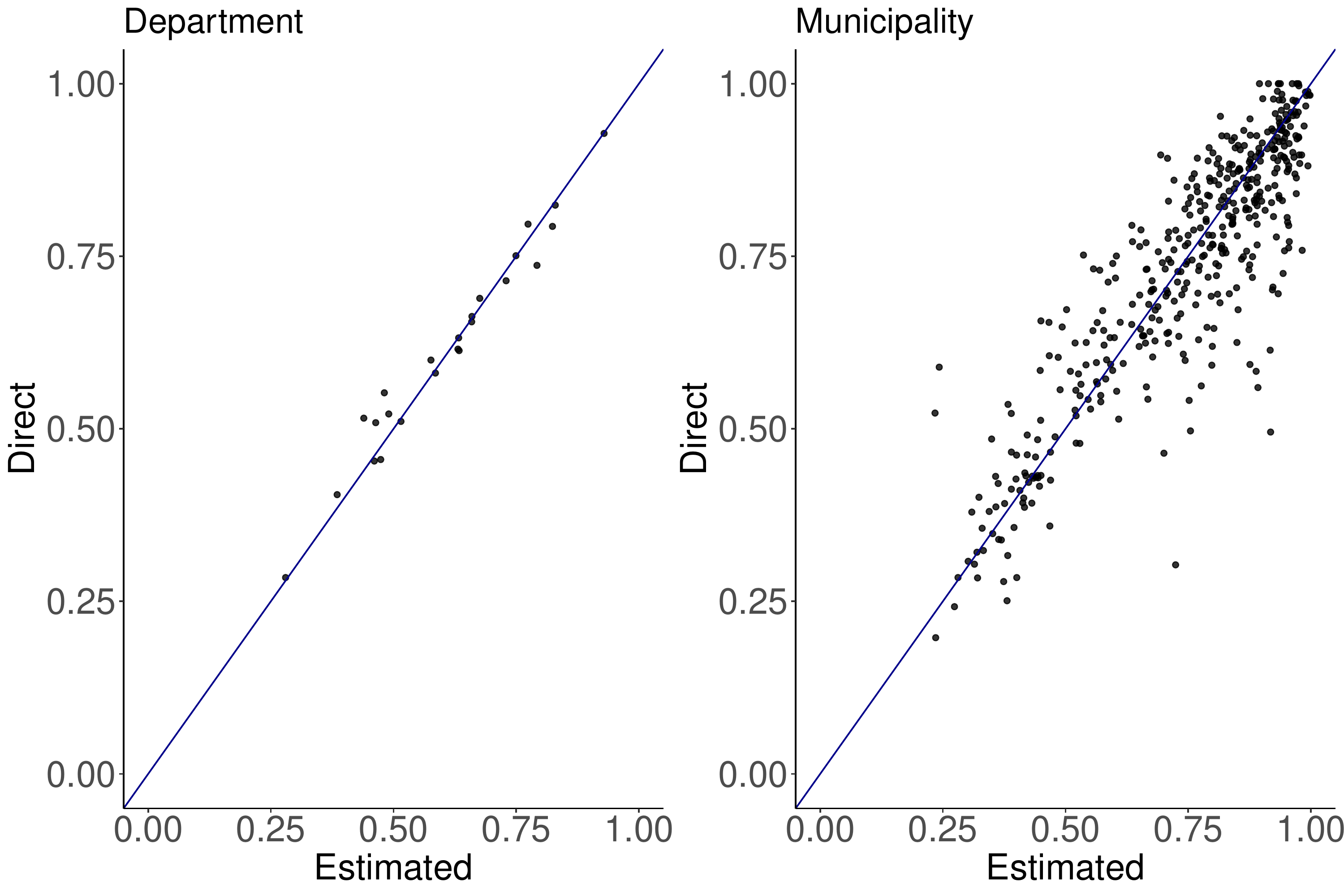}
\caption{Comparison between the direct and model-based estimates of the H at department and municipality level.} \label{fig:mdi}
\end{figure}

Direct and model-based estimates of in-sampled domains are compared in Figure \ref{fig:mdi}. As expected, for both cases, at the department and municipality level, the point estimates produced with the proposed method are very close to the direct estimates, with a Pearson correlation of 0.984 and 0.880 respectively. 

The performance of the estimator proposed in Section \ref{subsec:sae} is evaluated with a design-based simulation experiment.

To make the evaluation on a realistic case, we use the National population and housing census of Colombia 2018 as a fixed population and repeated samples are taken from it under a simple random sampling design with different sample sizes: a) 500, b) 5000, c) 50000 observations. A final scenario d) is performed considering a complex sampling design with 550000 observations, as in the original case study. The complex design mirrors the characteristics Great Integrated Household Survey (GEIH) of Colombia 2018 described in Subsection \ref{subsec:data}. 

Similarly as in the application, the target indicator is the multidimensional poverty index for small domains of Colombia (i.e., departments and municipalities). There are six of eight variables already available in the population data, which will be held as fixed. Two variables will be obtained following the methodology explained in Section \ref{subsec:mdi}. The performance of the small area predictors is evaluated with the coefficient of variation (previously described), the bias, and the root mean squared error for each domain. The last two measures are defined as:

$$\text{Bias}(\widehat{\text{H}}_d) = \frac{1}{T} \sum^{T}_{t=1} \Big(\widehat{\text{H}}_d^{(t)} - \text{H}_d \Big)\textcolor{white}{.};$$

$$\textcolor{white}{.....} \text{RMSE} (\widehat{\text{H}}_d)= \sqrt{\frac{1}{T} \sum^{T}_{t=1} \Big(\widehat{\text{H}}_d^{(t)} - \text{H}_d \Big)^2} \textcolor{white}{.},$$

where $\text{H}_d$ was defined in Section \ref{subsec:mdi}, and subscript $t$ indicates the $T$ simulations runs, in this exercise we set $T=100$. 

\begin{figure}[!htb]
     \centering
	\includegraphics[width=12cm]{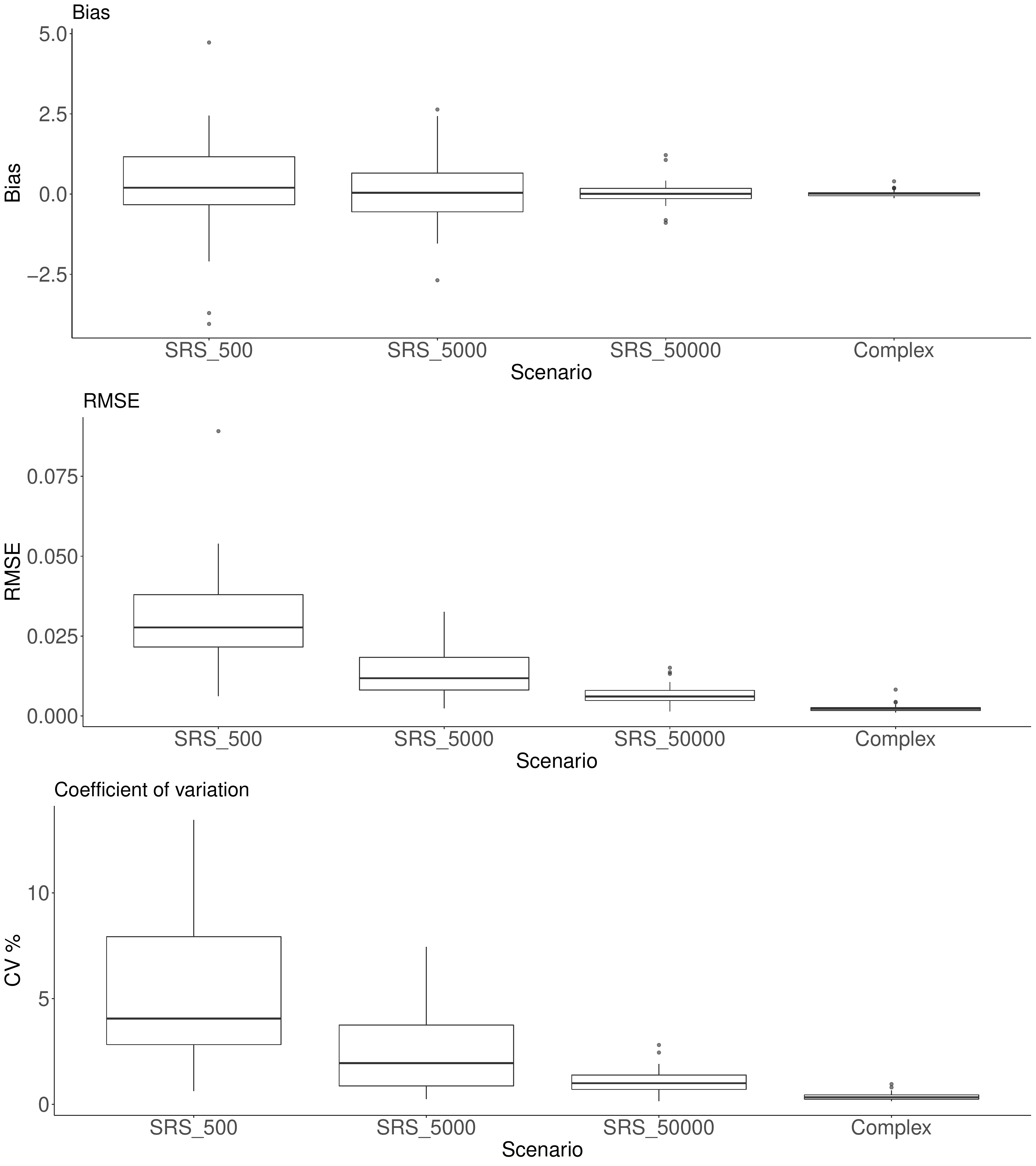}
 \caption{Performance measures of the area-specific point estimates for the multidimensional poverty incidence under different sample sizes and sampling designs.}
 \label{fig:evaluation}
 \end{figure}

Figure \ref{fig:evaluation} reports the distributions of the domain-specific bias, RMSE and CVs over domains for the evaluated estimator. As expected, increasing the sample size reduces the bias and increases the accuracy. This is clear for the ``complex" case, which has a larger sample size than the other three scenarios.

\section{Concluding Remarks and Further Research}\label{sec:conclusion}

Composite indicators are of great value in the study of complex phenomena and are widely used for public policies. Given the growing need for disaggregated information, there is a challenge to also increase the level of disaggregation of these types of indicators. Small area estimation methods address this problem, although there is no literature for the specific case of composite indicators. This paper aims to reduce this lack of information by proposing a methodology to obtain small area estimates when the indicator of interest is composed of dichotomous variables. We exemplify our approach with the incidence of multidimensional poverty (H). 

The challenge of producing small area estimates when working with composite indicators, such as the H, derives many questions that require further investigation. First,  a more general approach that includes the analysis of covariance structures and dependencies among indicators and dimensions might be considered. In this case study, exactly two indicators were estimated under the assumption that no dependencies between them (and other indicators) exist.  Second, the time gap between the sources of information (e.g. census and survey data) is another strain of research. Especially, if some of the indicators are available in census data and others would be estimated using up-to-date survey data. Third, regarding the point and MSE estimation, a model-based simulation could help to validate the proposed methodology. Fourth, benchmark procedures might be further investigated for correcting possible inconsistencies between the different estimates (e.g coming from the effects of sampling and non-sampling errors). Last, but not least, the final H, which is obtained by applying SAE methods, aims to capture the complexity of poverty along multiple dimensions of well-being – housing, water and sanitation, energy and connectivity, education, and employment
and social protection. However, the intra-household inequality and inequality amongst the poor population need to be captured in further research. Such measures, which can capture the big picture of poverty in a country at the most required disaggregated areas have become a critical underpinning for policy-relevant applications.

Further methods are needed to obtain small area estimates when the variables that make up the composite indicators are not dichotomous. A clear example is the Human Development Index (HDI) which is composed of life expectancy, years of education, and the Gini coefficient. This paper focuses on the estimation of one component of the global MPI. Further research is required to be able to compute the complete index, considering the intensity of poverty.

\vspace{1cm}

\textbf{Acknowledgments}
The authors gratefully acknowledge the support by UAEU Start-up Research Grant from the United Arab Emirates University.

\newpage

\bibliography{referencias}

\begin{thebibliography}{}

\bibitem[Alkire and Foster, 2007]{Alkire2007}
Alkire, S. and Foster, J. (2007).
\newblock Counting and multidimensional poverty measures.
\newblock {\em OPHI working paper 7}.
\newblock Available at: https://ophi.org.uk/working-paper-number-07/ (accessed
  June 2021).

\bibitem[Alkire and Santos, 2010]{alkire2010}
Alkire, S. and Santos, M.~E. (2010).
\newblock Acute multidimensional poverty: A new index for developing countries.
\newblock {\em Oxford Poverty \& Human Development Initiative (OPHI) Working
  Paper}.

\bibitem[Babenko et~al., 2017]{Babenko2017}
Babenko, B., Hersh, J., Newhouse, D., Ramakrishnan, A., and Swartz, T. (2017).
\newblock Poverty mapping using convolutional neural networks trained on high
  and medium resolution satellite images, with an application in mexico.

\bibitem[{CEPAL}, 2013]{CEPAL2013}
{CEPAL} (2013).
\newblock {\em Panorama Social de América Latina}.
\newblock {Santiago de Chile}.

\bibitem[{CEPAL}, 2014]{CEPAL2014}
{CEPAL} (2014).
\newblock {\em Panorama Social de América Latina}.
\newblock {Santiago de Chile}.

\bibitem[Chambers et~al., 2016]{chambers2016semiparametric}
Chambers, R., Salvati, N., and Tzavidis, N. (2016).
\newblock Semiparametric small area estimation for binary outcomes with
  application to unemployment estimation for local authorities in the uk.
\newblock {\em Journal of the Royal Statistical Society: Series A (Statistics
  in Society)}, 179(2):453--479.

\bibitem[Chandra et~al., 2018]{Chandra2018}
Chandra, H., Aditya, K., and Sud, U.~C. (2018).
\newblock Localised estimates and spatial mapping of poverty incidence in the
  state of bihar in india—an application of small area estimation techniques.
\newblock {\em PLOS ONE}, 13(6):1--14.

\bibitem[{DANE}, 2019]{DANE2019}
{DANE} (2019).
\newblock {Ficha Metodológica Censo Nacional de Población y Vivienda 2018}.
\newblock Technical report.

\bibitem[Deng and Wu, 2013]{Chengbin2013}
Deng, C. and Wu, C. (2013).
\newblock Improving small-area population estimation: An integrated geographic
  and demographic approach.
\newblock {\em Annals of the Association of American Geographers},
  103(5):1123--1141.

\bibitem[Engstrom et~al., 2022]{engstrom2022poverty}
Engstrom, R., Hersh, J., and Newhouse, D. (2022).
\newblock Poverty from space: Using high resolution satellite imagery for
  estimating economic well-being.
\newblock {\em The World Bank Economic Review}, 36(2):382--412.

\bibitem[Freudenberg, 2003]{freudenberg2003composite}
Freudenberg, M. (2003).
\newblock Composite indicators of country performance.

\bibitem[Gonz{\'a}lez-Manteiga et~al., 2007]{gonzalez2007estimation}
Gonz{\'a}lez-Manteiga, W., Lombard{\'\i}a, M.~J., Molina, I., Morales, D., and
  Santamar{\'\i}a, L. (2007).
\newblock Estimation of the mean squared error of predictors of small area
  linear parameters under a logistic mixed model.
\newblock {\em Computational statistics \& data analysis}, 51(5):2720--2733.

\bibitem[Gorelick et~al., 2017]{gorelick2017google}
Gorelick, N., Hancher, M., Dixon, M., Ilyushchenko, S., Thau, D., and Moore, R.
  (2017).
\newblock Google earth engine: Planetary-scale geospatial analysis for
  everyone.
\newblock {\em Remote Sensing of Environment}.

\bibitem[Harvey, 2002]{Harvey2002}
Harvey, J.~T. (2002).
\newblock Estimating census district populations from satellite imagery: Some
  approaches and limitations.
\newblock {\em International Journal of Remote Sensing}, 23(10):2071--2095.

\bibitem[Hobza and Morales, 2016]{hobza2016empirical}
Hobza, T. and Morales, D. (2016).
\newblock Empirical best prediction under unit-level logit mixed models.
\newblock {\em Journal of official statistics}, 32(3):661.

\bibitem[Jiang, 2003]{jiang2003empirical}
Jiang, J. (2003).
\newblock Empirical best prediction for small-area inference based on
  generalized linear mixed models.
\newblock {\em Journal of Statistical Planning and Inference},
  111(1-2):117--127.

\bibitem[Jiang and Lahiri, 2001]{jiang2001empirical}
Jiang, J. and Lahiri, P. (2001).
\newblock Empirical best prediction for small area inference with binary data.
\newblock {\em Annals of the Institute of Statistical Mathematics},
  53(2):217--243.

\bibitem[{Joint Research Centre-European Commission and OECD},
  2008]{joint2008handbook}
{Joint Research Centre-European Commission and OECD} (2008).
\newblock {\em Handbook on constructing composite indicators: methodology and
  user guide}.
\newblock OECD.

\bibitem[Koebe et~al., 2022]{koebe2022intercensal}
Koebe, T., Arias-Salazar, A., Rojas-Perilla, N., and Schmid, T. (2022).
\newblock Intercensal updating using structure-preserving methods and satellite
  imagery.
\newblock {\em Journal of the Royal Statistical Society: Series A (Statistics
  in Society)}.

\bibitem[Masaki et~al., 2020]{masaki2020small}
Masaki, T., Newhouse, D., Silwal, A.~R., Bedada, A., and Engstrom, R. (2020).
\newblock Small area estimation of non-monetary poverty with geospatial data.

\bibitem[Morales et~al., 2021]{morales2021course}
Morales, D., Esteban, M.~D., P{\'e}rez, A., and Hobza, T. (2021).
\newblock {\em A course on small area estimation and mixed models}.
\newblock Springer.

\bibitem[Moretti et~al., 2020]{moretti2020multivariate}
Moretti, A., Shlomo, N., and Sakshaug, J.~W. (2020).
\newblock Multivariate small area estimation of multidimensional latent
  economic well-being indicators.
\newblock {\em International Statistical Review}, 88(1):1--28.

\bibitem[Moretti et~al., 2021]{moretti2021small}
Moretti, A., Shlomo, N., and Sakshaug, J.~W. (2021).
\newblock Small area estimation of latent economic well-being.
\newblock {\em Sociological Methods \& Research}, 50(4):1660--1693.

\bibitem[{OECD}, 2023]{OECDCLI2023}
{OECD} (2023).
\newblock {Composite leading indicators}.
\newblock
  \url{https://www.oecd-ilibrary.org/economics/data/main-economic-indicators/composite-leading-indicators_data-00042-en}.
\newblock Accessed on 12 March 2023.

\bibitem[Pfeffermann, 2013]{Pfeffermann2013}
Pfeffermann, D. (2013).
\newblock New important developments in small area estimation.
\newblock {\em Statistical Science}, 28:40--68.

\bibitem[Pokhriyal and Jacques, 2017]{pokhriyal2017combining}
Pokhriyal, N. and Jacques, D.~C. (2017).
\newblock {Combining disparate data sources for improved poverty prediction and
  mapping}.
\newblock {\em Proceedings of the National Academy of Sciences},
  114(46):E9783--E9792.

\bibitem[Pratesi, 2016]{Pratesi2016}
Pratesi, M. (2016).
\newblock {\em Analysis of poverty data by small area estimation}.
\newblock John Wiley \& Sons.

\bibitem[Rao and Molina, 2015]{Rao2015}
Rao, J. N.~K. and Molina, I. (2015).
\newblock {\em Small area estimation}.
\newblock Wiley, New York, 2 edition.

\bibitem[Santos and Villatoro, 2018]{santos2018multidimensional}
Santos, M.~E. and Villatoro, P. (2018).
\newblock A multidimensional poverty index for latin america.
\newblock {\em Review of Income and Wealth}, 64(1):52--82.

\bibitem[Steinnocher et~al., 2019]{Steinnocher2019}
Steinnocher, K., Bono, A.~D., Chatenoux, B., Tiede, D., and Wendt, L. (2019).
\newblock Estimating urban population patterns from stereo-satellite imagery.
\newblock {\em European Journal of Remote Sensing}, 52(sup2):12--25.

\bibitem[{Transparency International}, 2023]{transparency2023}
{Transparency International} (2023).
\newblock {Corruption Perceptions Index}.
\newblock \url{https://www.transparency.org/en/cpi/2021}.
\newblock Accessed on 12 March 2023.

\bibitem[Tzavidis et~al., 2018]{tzavidis2018start}
Tzavidis, N., Zhang, L.-C., Luna~Hernandez, A., Schmid, T., and Rojas-Perilla,
  N. (2018).
\newblock From start to finish: a framework for the production of small area
  official statistics.
\newblock {\em Journal of the Royal Statistical Society: Series A (Statistics
  in Society)}, 181(4):927--979.

\bibitem[{UNDP}, 2023a]{UNDP23GDI}
{UNDP} (2023a).
\newblock {Gender Development Index}.
\newblock \url{https://hdr.undp.org/gender-development-index#/indicies/GDI}.
\newblock Accessed on 12 March 2023.

\bibitem[{UNDP}, 2023b]{UNDP23}
{UNDP} (2023b).
\newblock {Human Development Index}.
\newblock
  \url{https://hdr.undp.org/data-center/human-development-index#/indicies/HDI}.
\newblock Accessed on 12 March 2023.

\bibitem[{UNDP}, 2023c]{UNDP23MPI}
{UNDP} (2023c).
\newblock {Multidimensional Poverty Index}.
\newblock
  \url{https://hdr.undp.org/content/2022-global-multidimensional-poverty-index-mpi#/indicies/MPI}.
\newblock Accessed on 12 March 2023.

\bibitem[{United Nations General Assembly},
  2015]{UnitedNationsGeneralAssembly2015ResDevelopment}
{United Nations General Assembly} (2015).
\newblock {Res 70/1. Transforming Our World: The 2030 Agenda for Sustainable
  Development}.
\newblock Technical report, United Nations General Assembly.

\bibitem[Wolf et~al., 2022]{EPI2022}
Wolf, M.~J., Emerson, J.~W., C, E.~D., de~Sherbin, A., and A, W.~Z. (2022).
\newblock {\em 2022 Environmental Performance Index}.
\newblock {CT: Yale Center for Environmental Law and Policy}.

\end{thebibliography}

\appendix

\section{Proof when one indicator is missing}\label{proof:one}

\begin{proof}
We define $W=\alpha Y$, it is straightforward to find the probability mass function of $W$ as:
\begin{align*}
f_W\left(w\right)= & \begin{cases}
1-\pi& \text{if }w=0,\\
\pi & \text{if }w=\alpha,\\
0 & \text{otherwise}.
\end{cases}
\end{align*}

On the other hand: 
\begin{align*}
E\left(Z\right) & =1-P\left(X<\delta\right),\\
 & =1-P\left(\alpha Y+k<\delta\right),\\
 & =1-P\left(\alpha Y<\delta-k\right),\\
 & =1-P\left(W<\delta-k\right).
\end{align*}
The above expression depends on the value of $\delta-k$ as follows:
\begin{description}
\item [{Case~1}] When $\delta-k\le 0$,  $P\left(W<\delta-k\right)=0$.
\item [{Case~2}] When $0<\delta-k\le\alpha$,  
\[
P\left(W<\delta-k\right)=P\left(W=0\right)=1-\pi.
\]
\item [{Case~3}] When $\alpha<\delta-k$, therefore 
\begin{align*}
P\left(W<\delta-k\right) & =P\left(W=0\right)+P\left(W=\alpha\right)=1.
\end{align*}
\end{description}
This way, we conclude: 

\begin{align*}
E\left(Z\right)= & 
\begin{cases}
1 & \text{if }\delta-k\le0,\\
\pi & \text{if }0<\delta-k\le\alpha,\\
0 & \text{if }\alpha<\delta-k.
\end{cases}
\end{align*}
\end{proof}

\section{Proof when two indicators are missing}\label{proof:two}

\begin{proof}
We define $W=\alpha\left(Y_{1}+Y_{2}\right)$ a linear combination of the random variables $Y_{1}$ y $Y_{2}$. Note that the combination $Y_{1}+Y_{2}$ can only take the values of 0, 1 o 2. Therefore, $W$ can take the values of 0, $\alpha$ and
$2\alpha$, with the 
following probabilities: 

\begin{align*}
P\left(W=0\right)&=P\left(Y_{1}=0,Y_{2}=0\right)=\left(1-\pi_{1}\right)\left(1-\pi_{2}\right)\\
P\left(W=\alpha\right)&=P\left(Y_{1}=0,Y_{2}=1\right)+P\left(Y_{1}=0,Y_{2}=1\right)\\
&= \pi_{2}\left(1-\pi_{1}\right)+\pi_{1}\left(1-\pi_{2}\right)\\
P\left(W=2\alpha\right)&=P\left(Y_{1}=1,Y_{2}=1\right)=\pi_{1}\pi_{2}.
\end{align*}

The probability mass function of $W$ is then defined as:
\begin{align*}
f_W\left(w\right)= & \begin{cases}
\left(1-\pi_{1}\right)\left(1-\pi_{2}\right) & \text{if }w=0,\\
\pi_{2}\left(1-\pi_{1}\right)+\pi_{1}\left(1-\pi_{2}\right) & \text{if }w=\alpha,\\
\pi_{1}\pi_{2} & \text{if }w=2\alpha,\\
0 & \text{otherwise}.
\end{cases}
\end{align*}

On the other hand: 
\begin{align*}
E\left(Z\right) & =1-P\left(X<\delta\right),\\
 & =1-P\left(\alpha\left(Y_{1}+Y_{2}\right)+k<\delta\right),\\
 & =1-P\left(\alpha\left(Y_{1}+Y_{2}\right)<\delta-k\right),\\
 & =1-P\left(W<\delta-k\right).
\end{align*}
The above expression depends on the value of $\delta-k$ as follows:
\begin{description}
\item [{Case~1}] When $\delta-k\le 0$,  $P\left(W<\delta-k\right)=0$.
\item [{Case~2}] When $0<\delta-k\le\alpha$,  
\[
P\left(W<\delta-k\right)=P\left(W=0\right)=\left(1-\pi_{1}\right)\left(1-\pi_{2}\right).
\]
\item [{Case~3}] When $\alpha<\delta-k\le2\alpha$, therefore 
\begin{align*}
P\left(W<\delta-k\right) & =P\left(W=0\right)+P\left(W=\alpha\right)\\
 & =\left(1-\pi_{1}\right)\left(1-\pi_{2}\right)+\pi_{2}\left(1-\pi_{1}\right)+\pi_{1}\left(1-\pi_{2}\right).
\end{align*}
\item [{Case~4}] When $2\alpha<\delta-k$, therefore
\begin{align*}
P\left(W<\delta-k\right) & =P\left(W=0\right)+P\left(W=\alpha\right)+P\left(W=2\alpha\right)=1.
\end{align*}
\end{description}
By following the above cases we conclude: 

\begin{align*}
E\left(Z\right)= & 
\begin{cases}
1 & \text{if }\delta-k\le0,\\
\pi_{2}\left(1-\pi_{1}\right)+\pi_{1}\left(1-\pi_{2}\right)+\pi_{1}\pi_{2} & \text{if }0<\delta-k\le\alpha,\\
\pi_{1}\pi_{2} & \text{if } \alpha<\delta-k\le2\alpha,\\
0 & \text{if }2\alpha<\delta-k.
\end{cases}
\end{align*}
\end{proof}

\end{document}